# Testing Cosmic Homogeneity Using Galaxy Clusters

Michael J. Longo[1]

Department of Physics, University of Michigan, Ann Arbor, MI 48109-1040

According to the cosmological principle, galaxy cluster sizes and cluster densities, when averaged over sufficiently large volumes of space, are expected to be constant everywhere, except for a slow variation with look-back time (redshift). Thus, average cluster sizes or correlation lengths provide a means of testing for homogeneity that is almost free of selection biases. Using ~$10^6$ galaxies from the SDSS DR7 survey, I show that regions of space separated by ~2 Gpc/h have the same average cluster size and density to 5 – 10%. I show that the average cluster size, averaged over many galaxies, remains constant to <10% from small redshifts out to redshifts of 0.25. The evolution of the cluster sizes with increasing redshift gives fair agreement when the same analysis is applied to the Millennium Simulation. However, the MS does not replicate the increase in cluster amplitudes with redshift seen in the SDSS data. This increase is shown to be caused by the changing composition of the SDSS sample with increasing redshifts. There is no evidence to support a model that attributes the SN Ia dimming to our happening to live in a large, nearly spherical void.

## 1. INTRODUCTION

One of the most cherished tenets of modern cosmology is the cosmological principle that, on sufficiently large scales, the universe is homogeneous and isotropic. Until recently, quantitative tests of it were limited by the lack of deep astronomical surveys. One expectation is that the length scales and amplitudes associated with galaxy clustering are constant everywhere except for a slow variation with look-back time (or redshift). In this article I show that the average size of galaxy clusters and clustering amplitudes can be measured to a precision ~5% to test for homogeneity over distance scales ranging from 200 to 2000 Mpc/h. The SDSS galaxy distributions [1] are compared to those in the Millennium simulation [2].

Cluster sizes and counts are an important tool for understanding dark energy and other questions in cosmology. (See, for example, Ref. [3].) Techniques that require explicit cluster finding involve small samples and are also subject to serious redshift-dependent biases [4].

In this analysis, clusters are not identified explicitly. Instead, I use galaxy separation correlation lengths and amplitudes as proxies for cluster sizes and spatial densities. For simplicity, I refer to these statistical correlations as "clusters". The technique described here does not require cluster finding, and it appears to be almost bias free. The only assumption is that when averaged over sufficiently large volumes of space the clustering length of galaxies is constant or evolves slowly with redshift (i.e., look-back time) for redshifts $z \lesssim 1$. The very large SDSS data set allows a detailed study of systematic effects without the need to resort to simulations to correct for biases. Selection effects in the galaxy data do change significantly with increasing $z$; however, this does not affect the comparison of cluster sizes and densities at the same $z$.

---

[1] mlongo@umich.edu

## 2. THE SDSS SAMPLE

The SDSS DR7 database [1] contains ~800,000 galaxies with spectra for $z \lesssim 1$. Their spectroscopic redshifts have a typical uncertainty ~0.1%. The sky coverage of the spectroscopic data is almost complete for right ascensions ($\alpha$) between 110° and 240° and declinations ($\delta$) between –5° and 60°. For the hemisphere toward $\alpha = 0°$, only narrow bands in $\delta$ near –10°, 0°, and +10° are covered. The following discussion will be limited to the hemisphere toward $\alpha = 180°$ except as noted. All objects with spectra that were classified as "galaxies" in the SDSS DR7 database were used in this analysis.

## 3. THE ANALYSIS

The galaxies were binned in bins of redshift, right ascension, and declination. The 3-dimensional comoving distance between every pair of galaxies in each $z$ bin was calculated. For an $\Omega = 1$ universe the comoving distance between two closely spaced objects is given by the same expression as in Euclidean space. In Cartesian coordinates, in terms of right ascension, declination, and comoving line-of-sight distance $r(z)$, the coordinates are

$$x' = r_p \cos\alpha, \quad y' = r_p \sin\alpha, \quad z' = r \sin\delta \quad \text{where} \quad r_p = r \cos\delta \qquad (1)$$

This gives a right-handed coordinate system with the $x'$-axis along $\alpha = 0°$, the $y'$-axis along $\alpha = 90°$, and the $z'$-axis along $\delta = 90°$. The connection between the redshift and the comoving line-of-sight distance $r(z)$ is determined by the cosmological model chosen. Since $r$ is just a scale factor that determines the overall size of clusters, the choice of a model is not critical to this analysis. I use a flat CDM model with $\Omega_m = 0.3$, $\Omega_\Lambda = 0.7$. The luminosity distances were calculated using the IDL subroutine LUMDIST [5]. The comoving separation $d$ between two nearby objects is then

$$d = \sqrt{\delta x'^2 + \delta y'^2 + \delta z'^2} \qquad (2)$$

where $\delta x', \delta y', \delta z'$ are the finite differences along the $x'$, $y'$, $z'$ axes respectively.

The galaxies were first binned in redshift slices between 0.02 and 1.0. Three-dimensional separations for every pair of galaxies in the redshift bin were calculated and binned in 125 bins between 0 and 0.010 in dimensionless units (approx. 0 to 30 Mpc). The separations were then calculated with the galaxies' right ascensions, declinations, and redshifts randomly scrambled among the galaxies in that redshift bin. To reduce the statistical fluctuations in the scrambling process, 10 scrambling runs were averaged. Figure 1 compares the ratio of the not-scrambled (NS) and scrambled (S) distributions for the number of pairs in each bin vs. separation $d$ for typical $z$ ranges. The peaking at small separations is a clear signal for galaxy clustering on a length scale ~0.003 in redshift units (~9 Mpc). The width of the peak is a measure of the average size of clusters in the volume studied, and the amplitude of the peak or its integral measures of the clustering strength.

Occasionally in the SDSS data, nearby galaxies appear more than once with different IDs when different points in the same galaxy are chosen as centers. Therefore pairs with separations $<1\times10^{-5}$ were excluded in order to remove possible duplicates. (This is several times the radius of a typical large galaxy.) Because the numbers of galaxy pairs were slightly different for the NS and S samples, the ratios were normalized to an average of 1.0 in the separation range 0.0064 to 0.010.

The solid curves in Fig. 1 are fits to an exponential, $R-1 = a_0 e^{-d/a_1}$, where $R$ is the *NS/S* ratio, $a_0$ is the amplitude of the exponential, and $a_1$ is a measure of the width of the distribution. The integral of the exponential is $a_0 a_1$. A least-squares fit to the exponential using bins 2 through 60, corresponding to $0.000125 < d < 0.0048$, was made. The first bin was not used in order to avoid systematic effects due to the finite resolution of the SDSS camera (e.g., "fiber collisions"). The exponential fit, though simple and convenient, is not physically motivated, and sometimes gave poor fits or failed completely. Therefore, a numerical measure of the rms width of the peak $W_{sum}$ was determined from the contents of the bins, excluding the first bin. As a measure of the clustering strength, $I_{sum}$ is defined as a numerical integral (sum) of the contents of the bins, excluding the first. Generally the widths and integrals from the exponential fit tracked those from the histogram bins very well. However, those from the contents of the bins typically had somewhat smaller uncertainties and were more robust when the bin occupancy was small. In the following, only the widths and integrals from the sum over bins will be given.

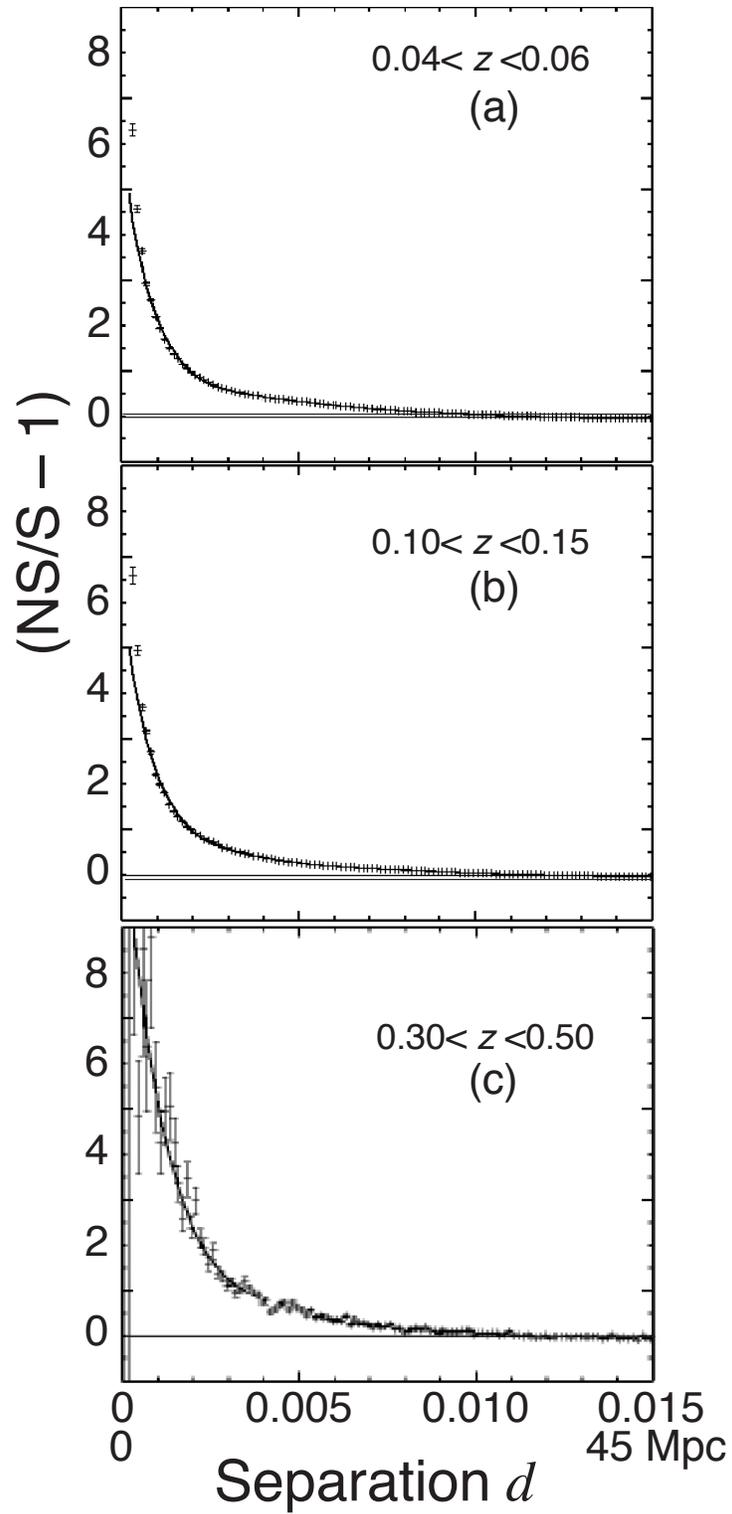

Figure 1 – Typical plots of *NotScrambled/Scrambled* correlation vs. separation in dimensionless units and Mpc for 3 redshift ranges. The smooth curves are a fit to an exponential.

The uncertainties in the fitted widths and integrals were estimated in several ways. Statistical errors could be estimated from the contents of the separation bins with the correlations between the points taken into account. A resampling method using 10 non-overlapping subsamples was also used to estimate the uncertainties. The resampling method gave the most plausible errors; however, it failed for the large $z$ ranges where the number of galaxies in each separation bin was small. The resampling method typically gave uncertainties ~50% larger than the statistical ones. In the following, to be conservative, the resampling uncertainties were used where practical.

The effects of systematic uncertainties, such as edge effects, depend on the context. Galaxies at the edge of the defined volumes will have neighbors on one side only. These effects are expected to be small as long as the dimensions of the slice are much larger than the cluster size or 9 Mpc. Edge effects were minimized by using the same size angle bins when looking for possible variations in the widths and integrals with right ascension and declination. Tests for this were made by dividing the slices into smaller ranges of right ascension and declination. No significant changes in the parameters were seen for $z > 0.04$. Regions near the limits of the SDSS coverage were also not used. The redshift dependence is more complicated, particularly because the spatial density of observed galaxies decreases with increasing $z$ since only the brightest galaxies are above the detection threshold at higher $z$. Higher $z$ also corresponds to earlier epochs when the clusters were more diffuse (i.e., larger comoving widths and lower amplitudes or integrals). Thus, the only way to test the $z$ dependence is to compare that observed in the SDSS galaxies with that expected from a simulation. This is discussed in Sections 5 and 6.

## 4. VARIATION OF WIDTHS AND INTEGRALS WITH $\alpha$, $\delta$ AND GALAXY DENSITY

Figure 2 compares the correlation widths and integrals for 5 right ascension bins, each 24° wide for 3 redshift ranges. Declinations between 0° and 60° were used. Error bars that are smaller than the symbol sizes are not shown. There is no significant variation of $W_{sum}$ or $I_{sum}$ with $\alpha$. Note that the $I_{sum}$ show a significant increase with increasing $z$. Similar plots (not shown) of these quantities vs. $\delta$ show no significant variation with $\delta$ in the range 0° to 60°.

The variation of $W_{sum}$ and $I_{sum}$ with galaxy number density could be studied by choosing a fraction of the SDSS galaxies at random. With ½ of the galaxies, $W_{sum}$ increased by about 4% and $I_{sum}$ decreased 6.5% for $0.04 < z < 0.06$. For $0.3 < z < 0.7$, $W_{sum}$ increased by about 22% and $I_{sum}$ decreased 35%.

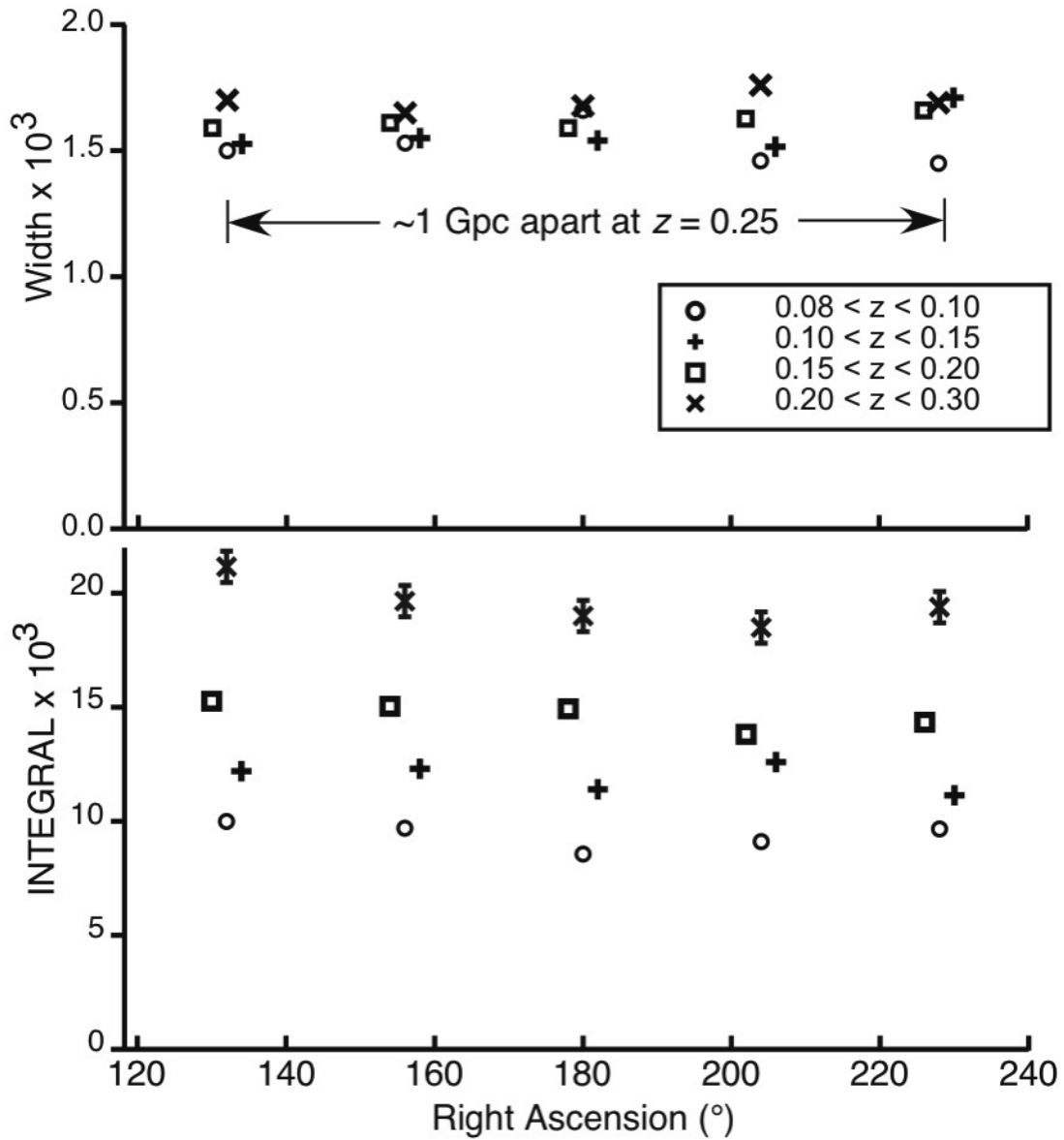

Figure 2 – Correlation widths and integrals vs. right ascension for 4 redshift ranges. Except where shown, the uncertainties are smaller than the symbols. These show no significant variation with angle. The integrals show a significant evolution with increasing redshift.

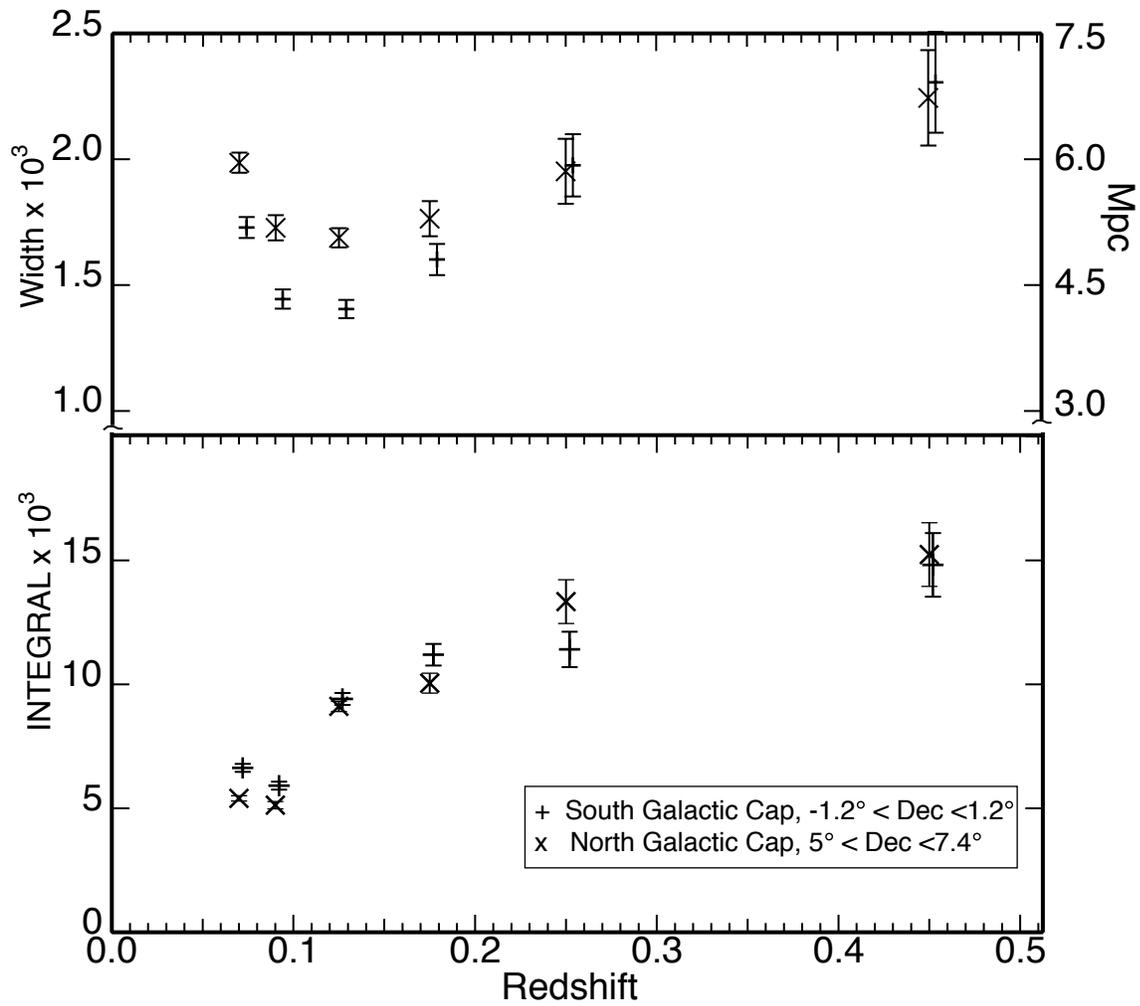

Figure 3 – Correlation widths and integrals vs. redshift for two regions with similar geometry and spatial density, one in the NGC and one in the SGC. For $z \sim 0.4$, these regions are separated by ~2.4 Gpc.

## 5. COMPARISON OF THE NORTH AND SOUTH GALACTIC CAP REGIONS

It is of interest to compare the correlation parameters over regions of space as far apart as possible. Most of the SDSS galaxies are in the north galactic cap region toward $(\alpha, \delta) = (193°, 27°)$. For the hemisphere toward $\alpha = 0°$, only narrow bands in $\delta$ near $-10°$, $0°$, and $+10°$ are covered. This disparity complicates a comparison of the NGC and SGC regions. To allow this comparison, two slices with the same width in $\delta$ and in opposite hemispheres were chosen. Both slices included a range of about 100° in $\alpha$. When necessary, the appropriate number of galaxies in the NGC was randomly selected so that the spatial densities of galaxies in both hemispheres were approximately equal. The $W_{sum}$ and $I_{sum}$ for 6 ranges of redshift are compared in Fig. 3. At the higher redshifts the $W_{sum}$ and $I_{sum}$ are in good agreement for the two hemispheres. However, for $z \lesssim 0.2$ there is some evidence of larger $W_{sum}$ and slightly smaller $I_{sum}$ toward the NGC, but it is likely that this small discrepancy may be due to edge effects. The regions studied are only 2.4° wide in declination and the volumes are quite small at the lower redshifts. Despite the attempt to match the geometries and spatial density in the two regions, the discrepancy may be due to systematic effects. On the other hand, the good agreement for $z > 0.2$ shows that the correlation widths and amplitudes are very similar for regions of space that are ~2.4 Gpc away from each other.

## 6. VARIATION OF WIDTHS AND INTEGRALS WITH REDSHIFT

Figure 4 shows variation of the correlation widths and integrals with redshift. The circles are for the SDSS galaxies, and the triangles are for the Millennium simulation that will be discussed in the next section. The statistical errors are usually smaller than the symbols. However, there are relatively large uncertainties in comparing the data with the simulation, especially for larger redshifts, because the simulation does not reproduce the selection biases in the SDSS data. However, the trend of the SDSS widths and integrals with increasing redshift is quite surprising. The SDSS only "sees" galaxies with green magnitudes up to about 20. Therefore at higher redshift the spatial density of galaxies decreases rapidly. As discussed above, this should lead to a significant <u>decrease</u> in the correlation integrals and a small increase in the widths, as the correlations are diluted. This general behavior is indeed shown by the Millennium simulation. Other possible explanations of this effect are discussed below in Sect. 8.

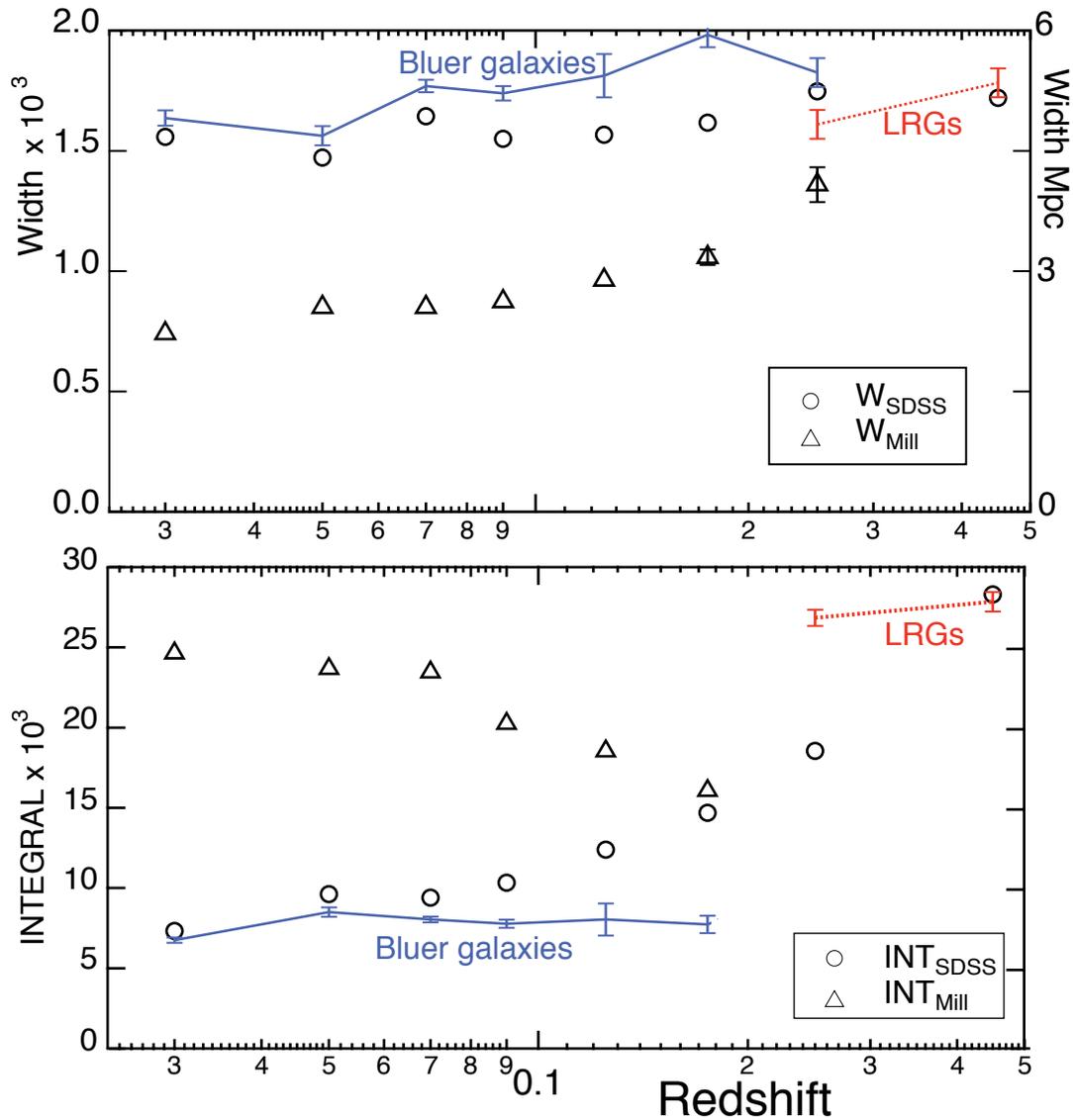

Figure 4 – Correlation widths and integrals vs. redshift. Except where shown, the uncertainties are smaller than the symbols. The circles are for the SDSS galaxies; the triangles are for the Millennium simulation. The dotted (red) and solid (blue) curves are for the redder and bluer SDSS galaxies respectively.

## 7. COMPARISON WITH THE MILLENNIUM SIMULATION

The Millennium Simulation (MS) is a very large simulation of the concordance ΛCDM cosmology [2]. This simulation follows the growth of dark matter structure from redshift $z = 127$ to the present using $10^{10}$ particles within a comoving box 500 $h^{-1}$Mpc on a side. In Croton et al. [6], semi-analytic models are applied to the output of the Millennium Run to simulate the growth of galaxies and their central supermassive black holes and to follow the detailed assembly history of each object. Public databases with the results are available at http://www.g-vo.org. The DeLucia2006a and DeLucia2006a_SDSS2MASS tables [7] give positions and other data for ~$10^7$ simulated galaxies with "snapshots" in time ranging from $z$=127 to the present with filter magnitudes mimicking the SDSS. I refer to the snapshots as epochs $z_E$ to distinguish them from SDSS redshifts that correspond to distances as well as look-back times. For each redshift, the appropriate MS epoch was used in determining the $W_{sum}$ and $I_{sum}$.

The analysis of the MS galaxies was done with the same programs used for the SDSS. The redshift, right ascension, and declination ranges were the same as those used for the SDSS. The SDSS galaxy redshifts have typical uncertainties ~4 x $10^{-5}$. This leads to a small smearing of the galaxy separations, which would lead to a small increase in the $W_{sum}$ and a decrease in the $I_{sum}$. The MS does not include this smearing so the MS $z$'s were smeared with a Gaussian distribution with the same width as that for the SDSS galaxies.

The 500 Mpc/h cube in the simulation only allowed redshift slices out to about $z$=0.2 to be studied. This could be extended somewhat by scaling the radial coordinate $r$ appropriately. Since the simulations were generated within a comoving box at a fixed look-back time, the redshift/distances $z$ were calculated as $r \equiv z c / H_0 = \sqrt{x'^2 + y'^2 + z'^2}$, where $x'$, $y'$, $z'$ are the coordinates relative to the origin chosen along one edge of the box. The separations were then calculated as $d = \sqrt{\delta x'^2 + \delta y'^2 + \delta z'^2}$ as before. For each redshift bin a subset of the MS galaxies was chosen at random to give approximately the same number of galaxies in each redshift slice as in the SDSS. The resulting $W_{sum}$ and $I_{sum}$ are plotted vs. $z$ in Fig. 4 as the triangles.

The widths from the simulation appear to increase slowly with redshift. This is consistent with expectation as the spatial density of galaxies decreases with $z$. The SDSS widths, on the other hand, show little variation with $z$. The MS widths are considerably smaller, which isn't surprising since the simulation does not include effects such as galaxy collisions and mergers.

For small $z$, the MS $I_{sum}$ are much larger than the SDSS. Again this is probably due to the neglect of galaxy collisions and mergers. The SDSS $I_{sum}$ appear to be increasing with redshift, while the MS amplitudes are decreasing. Since the widths are almost constant, the $I_{sum}$ are a good proxy for galaxy number density and mass density. The analysis of both samples was done with the same program, the galactic densities were forced to be the same by choosing MS galaxies randomly, and the geometry of the MS slices were the same as for the SDSS, so that this dis-crepancy in their $z$ dependence is at first glance surprising.

A significant limitation in comparing the MS and SDSS correlations is in the MS treatment of "galaxy" luminosity. In the MS, the galaxy masses are used as a proxy for luminosity; the assigned magnitudes are thus artificial and they are strongly correlated with mass, unlike the situation with physical galaxies. The MS $W_{sum}$ and $I_{sum}$ in Fig. 4 were calculated using the entire luminosity range. If, for example, the MS sample for $0.15 < z < 0.20$ is divided into the brighter half and dimmer half, $W_{sum}$ is about 27% smaller and $I_{sum}$ is ~60% larger for the brighter half, while for the SDSS $W_{sum}$ and $I_{sum}$ both show no significant change between the two halves. Taking this effect into account in the MS would thus make it more consistent with the observations.

## 8. DISCUSSION

The correlation lengths found here, ≈4.5 Mpc/h, are roughly consistent with other measurements (Zehavi et al. [8]; Davis & Peebles [9]; Carlberg et al. [10]), though the sizes are definition dependent. There is no evidence for a significant change in cluster sizes with redshift out to $z$ =0.5. In contrast, Zehavi et al., using the standard $\xi(r) = (r/r_0)^{-\gamma}$ fit to SDSS data, find that $r_0$ increases from 2.83±0.19 Mpc/h for R-band luminosity $M_r \approx -17.5$ and $z \approx 0.02$ to 10.00±0.29 Mpc/h for $M_r \approx -22.5$ and $z \approx 0.17$. Carlberg et al., analyzing the CNOC2 high-luminosity sample with a luminosity-compensated absolute magnitude, find that $r_0$ <u>decreases</u> from 4.75±0.05 Mpc/h for $z \approx 0.10$ to 4.26±0.18 Mpc/h for $z \approx 0.49$.

This cluster size is also generally consistent with the measured value [11] of 0.8 for the cosmological parameter $\sigma_8$, as a box of side 8 Mpc/h contains about the amount of material to form a cluster.

There has been considerable discussion of alternatives to dark energy that attempt to explain the cosmic microwave background (CMB) data and the Type Ia supernovae data using an inhomogeneous universe. One alternative model supposes that we are located near the center of a large, underdense, nearly spherical void. (See, for example, Clifton, Ferreira, and Land [12].). By tuning the radial void profile, it is possible to match the luminosity distance-redshift relation of the concordance ΛCDM model (Yoo, Kai, & Nakao [13], Zibin, Moss, and Scott [14]). Zibin, et al. show that a void profile with the local matter density increasing by a factor of 5 between $z$=0.3 and $z$=1.0 is required to fit the CMB and supernova data.[2] Figure 4 shows a striking increase of the SDSS correlation amplitudes with redshift. It is tempting to try to attribute this discrepancy to an increase in galaxy density with increasing $z$ that is suggestive of our location near the center of a void.

---

[2] In a later paper [15] they show this model is inconsistent with the observed local Hubble rate.

However, a systematic effect we must consider is the change in composition of the SDSS sample with redshift. At larger $z$, the sample is dominated by luminous red galaxies (LRG) [16]. These might have different clustering properties than the fainter, bluer galaxies that are more numerous in the SDSS sample at smaller redshifts. A simple but useful technique to study the LRG sample is to use the filter magnitudes [17]. Here I use the distribution in ($u^*- z^*$), the difference in the filter magnitudes in the ultraviolet and far infrared. When this is plotted for each redshift band, there is a clear two-peak distribution for the larger redshifts with the LRG's at larger ($u^*- z^*$). For $z \gtrsim 0.2$, there are few of the bluer and fainter galaxies and the LRG's dominate the sample. For $z \lesssim 0.1$, the two peaks merge and there is no clear separation. Below $z \sim 0.2$ the sample is a mixture with the more numerous blue galaxies dominating at the lowest redshifts. Thus it is possible to study the clustering of a clean sample of LRG's for $z>0.2$. These regimes are shown in Fig. 4 as the red (dotted) curve and the blue (solid) curves. The LRG's show a slighter smaller $W_{sum}$ that suggests a somewhat more compact structure. The SDSS $I_{sum}$ are almost independent of $z$ for the bluer sample, while those for the LRG's are over 3 times larger. This shows the LRG's cluster much more strongly than the bluer galaxies. A more quantitative estimate of this effect can be obtained by increasing the galaxy number density. To produce this 3-fold increase in the $I_{sum}$ would require an increase by over a factor of 4 above that actually observed in the number density of SDSS galaxies at $z \sim 0.3$. If we restrict our attention to the "bluer" galaxies in Fig. 4, there is no evidence for an increase in clustering amplitude with increasing $z$. Thus, the apparent rise in $I_{sum}$ can be explained by the changing composition of the SDSS galaxies as the LRG fraction increases with redshift.

## 8. CONCLUSIONS

The concordance $\Lambda$CDM model of the Universe assumes the cosmological principle that the Universe is homogeneous and isotropic on sufficiently large scales. The cluster widths and amplitudes from the SDSS data provide a direct test for inhomogeneity. The graphs in Fig. 2 show that the cluster widths and amplitudes are constant to about 5% over distance scales ~ 1 Gpc. The comparison of the north galactic cap region with the south cap in Fig. 3 shows that they are equal to about 10% for regions separated by ~2.4 Gpc in the other direction.

I find that cluster sizes (or correlation lengths) remain approximately constant with redshift at about 4.8 Mpc/h out to $z \approx 0.5$. The clustering amplitudes (or integrals) show a large increase with increasing redshift, an effect that the Millennium Simulation does not emulate. An analysis shows this increase is due to an increasing fraction of the brighter LRG's at larger $z$. These appear to cluster more strongly, with clustering amplitudes over 3 times greater than those for the less luminous and bluer galaxies that dominate the sample at smaller redshifts. There is no need to invoke a large rise in galaxy number densities at $z \sim 0.4$ to explain this. In any case, this rise in cluster amplitudes at larger $z$ appears to be consistent in all directions, so that the void explanation is only tenable if we happen to be near the center of a large void that has a galaxy density $\sim 1/4^{th}$ that of the surrounding universe.

This analysis would not have been possible without the dedicated efforts of the SDSS collaboration. The Millennium Run simulation was carried out by the Virgo Supercomputing Consortium at the Computing Centre of the Max-Planck Society in Garching. The semi-analytic galaxy catalog is available at http://www.mpa garching.mpg.de/galform/agnpaper. I am grateful to G. Lemson for help in using the MS and to A. Evrard for helpful discussions of galaxy clustering.